\documentclass[sn-basic]{sn-jnl}

\usepackage{graphicx}%
\usepackage{multirow}%
\usepackage{amsmath,amssymb,amsfonts}%
\usepackage{amsthm}%
\usepackage{mathrsfs}%
\usepackage[title]{appendix}%
\usepackage{xcolor}%
\usepackage{textcomp}%
\usepackage{manyfoot}%
\usepackage{booktabs}%
\usepackage{algorithm}%
\usepackage{algorithmicx}%
\usepackage{algpseudocode}%
\usepackage{listings}%

\theoremstyle{thmstyleone}%
\theoremstyle{thmstyletwo}%

\theoremstyle{thmstylethree}%

\begin{document}


\title[Wall slip and flow heterogeneity in a sludge]{Wall slip and bulk flow heterogeneity in a sludge under shear}


\author*[1,2]{\fnm{Sébastien} \sur{Castel}}\email{sebastien.castel2@cea.fr}

\author[1]{\fnm{Arnaud} \sur{Poulesquen}}\email{arnaud.poulesquen@cea.fr}

\author[2,3]{\fnm{Sébastien} \sur{Manneville}}\email{sebastien.manneville@ens-lyon.fr}

\affil*[1]{\orgdiv{CEA, DES/ISEC/DPME/SEME/LFCM}, \orgname{Université de Montpellier}, \orgaddress{\street{Marcoule}, \postcode{F-30207} \city{Bagnols-sur-Cèze}, \country{France}}}

\affil[2]{\orgdiv{ENSL, CNRS, Laboratoire de Physique}, \postcode{F-69342} \city{Lyon}, \country{France}}

\affil[3]{\orgdiv{Institut Universitaire de France}}

\abstract{We investigate the shear flow of a sludge mimicking slurries produced by the nuclear industry and constituted of a dispersion of non-Brownian particles into an attractive colloidal dispersion at a total solid volume fraction of about 10 \%. Combining rheometry and ultrasound flow imaging, we show that, upon decreasing the shear rate, the flow transitions from a homogeneous shear profile in the bulk to a fully arrested plug-like state with total wall slip, through an oscillatory regime where strong fluctuations of the slip velocity propagate along the vorticity direction. When the shear stress is imposed close to the yield stress, the shear rate presents large, quasi-periodic peaks, associated with the propagation of local stick-and-slip events along the vorticity direction. Such complex dynamics, reminiscent of similar phenomena reported in much denser suspensions, highlight the importance of local flow characterization to fully understand sludge rheology.}




\maketitle

\section{Introduction}\label{sec1}

Sludges and slurries are ubiquitous not only in the industry, from mining, construction, and chemical reactors \cite{Tay:1997, Rakotonimaro:2017,Wang:2007} to wastewater treatment and soil remediation \cite{Robles:2008,Eshtiaghi:2013,Gherghel:2019} or energy storage \cite{Kraytsberg:2016,Hawley:2019}, but also in natural debris flows and landslides \cite{Enos:1977,Takahashi:1981,Hungr:1995}. Such complex fluids consist of suspensions of solid particles characterized by a large polydispersity, with sizes spanning both the colloidal domain ($\lesssim 1~\mu$m) and the non-Brownian range, up to very large grains ($\gtrsim 1~$mm). The interactions between suspended particles include electrostatic and van der Waals interactions, viscous and lubrication forces, and frictional contacts. Such interparticle forces are mediated by the suspending liquid, which generically contains a number of dissolved species such as ions or macromolecules, leading to complex solvation effects \cite{Israelachvili:2011,Coe:2023} or to depletion forces \cite{Mao:1995,Lekkerkerker:2024}. Thus, the composition of the suspending liquid has a key impact on the structure of sludges and slurries and on their flow behavior.

Overall, depending on the solid volume fraction, the rheological properties of sludges and slurries can be seen as resulting from an intricate combination of the physics of various ``model'' systems, namely hard-sphere suspensions, colloidal gels, colloidal glasses, and granular suspensions \cite{Coussot:1999,Roussel:2010,Mueller:2010,Lu:2013,Joshi:2014,Guazzelli:2018,Ness:2022}. This makes it very difficult to reach a general modelling of the flow behavior of sludges and slurries. More precisely, even the simplest, plane shear deformation leads to peculiar flow phenomena that strongly deviate from the usual shear-thinning or yield stress responses observed in model systems, and that question the very assumptions underlying the interpretation of standard rheological data \cite{Balmforth:2014,Bonn:2017}. These phenomena include, but are not restricted to, heterogeneous flows with apparent slippage at the walls \cite{Cloitre:2017} and/or shear localization within shear bands \cite{Ragouilliaux:2006,Ovarlez:2009,Divoux:2016a}, shear-induced particle migration \cite{Leighton:1987,Fall:2015}, time-dependence such as aging and thixotropy \cite{Baudez:2008}, and fluctuating or even chaotic dynamics in dense, shear-thickening slurries \cite{Lootens:2003,Nagahiro:2013,Hermes:2016,Rathee:2017,Bossis:2017,SaintMichel:2018,Ovarlez:2020}.

In this context, it is clear that rheometry alone cannot provide a full picture of the behavior of sludges under flow and must be complemented by local measurements, e.g. flow profilometry or spatially resolved measurements of the volume fraction. One major issue raised by sludges and slurries is the fact that they are optically opaque, therefore preventing the use of particle imaging velocimetry, confocal microscopy, or standard light scattering techniques \cite{Manneville:2008}. In recent years, a number of techniques have been coupled to rheometry in order to circumvent this difficulty and investigate ``non-model,'' opaque colloidal suspensions and granular pastes under shear. These combined techniques include magnetic resonance imaging \cite{Huang:2005,Ragouilliaux:2006,Ovarlez:2006}, ultrasound imaging \cite{Gallot:2013,SaintMichel:2016,Liberto:2020}, small-angle X-ray or neutron scattering \cite{Hipp:2019,Sudreau:2023}, X-ray radiography \cite{Gholami:2018,SaintMichel:2019}, X-ray tomography \cite{Yuan:2024}, and optical coherence tomography \cite{Harvey:2011}.

Here, we use ultrasound imaging to perform an in-depth characterization of the complex flow behavior of a surrogate sludge mimicking nuclear sludges. By imposing the shear rate in a small-gap concentric-cylinder geometry, we first show that this sludge displays strong wall slip all along the flow curve, in spite of the significant surface roughness of the shear cell. Upon decreasing the shear rate, the flow transitions from a homogeneous shear profile in the bulk to a fully arrested plug-like state through an oscillatory regime. In this intermediate range of shear rates, translational invariance is broken due to strong fluctuations of the slip velocity at the moving wall, which propagate along the vorticity direction. Second, we impose the shear stress in order to investigate the vicinity of the yield stress in more details. Here again, we uncover a remarkable unsteady flow regime where the shear rate presents large, quasi-periodic peaks, hinting at bistability. Flow imaging shows that these peaks are associated with the propagation of local stick-and-slip events along the vorticity direction. 

Our paper is structured as follows. Section~\ref{sec2} introduces the sludge under study, the rheological protocol, and the ultrasonic measurements, as well as the data analysis used in this work. Section~\ref{sec3} focuses successively on experiments performed under an imposed shear rate and under an imposed shear stress. Finally, our findings are discussed in light of recent results on more ``model'' systems in Section~\ref{sec4}. Our results confirm that combining rheometry and local velocimetry is crucial both for the fundamental understanding of sludge rheology and for the prediction of their flow properties.

\section{Materials and methods}\label{sec2}

\subsection{Composition of the surrogate sludge}\label{subsec21}

In the nuclear industry, coprecipitation sludges are produced through the addition of salts into a liquid effluent containing radioelements such as strontium, cesium, and certain actinides \cite{Pacary:2010,Flouret:2012,Yaghy:2023}. The added salts can form insoluble compounds when mixed together in the liquid effluent. They are chosen for presenting  chemical affinities with specific radioelements, which allows radioactive atoms to be dragged into the formatting insoluble compounds and eventually trapped inside the particle structure. After settling and filtration, a pasty, radioactive sludge is obtained. Here, we focus on a non-radioactive sludge produced at a pilot scale using a large mixer and a rotary filter pre-coated with  perlite, a siliceous sand used as a filter aid  \cite{Yang:2022}. The various constituents of the present sludge were selected in order to obtain a composition close to that of a coprecipitation sludge, yet without any radioactive element. Based on the very small fraction of radioelements in nuclear sludges, which is less than one part per million, we shall assume that their rheological behavior is only marginally affected by the presence of radioelements, and that the properties reported below on the surrogate sludge are representative of those of actual nuclear sludges. 

\begin{table}[htb]
\caption{Macroscopic properties of the sludge under study}\label{tab:prop}%
\begin{tabular}{@{}llll@{}}
\toprule
Property & Value \\
\midrule
Global density of the sludge   & $1.24 \text{ g.cm}^{-3}$ \\
Solid volume fraction    & 9.5 \%  \\
Ionic strength of the suspending liquid    & 0.5 mol.L$^{-1}$  \\
pH of the suspending liquid    & 6.5   \\
\botrule
\end{tabular}
\end{table}

\begin{figure}[t]%
\centering
\includegraphics[width=0.9\textwidth]{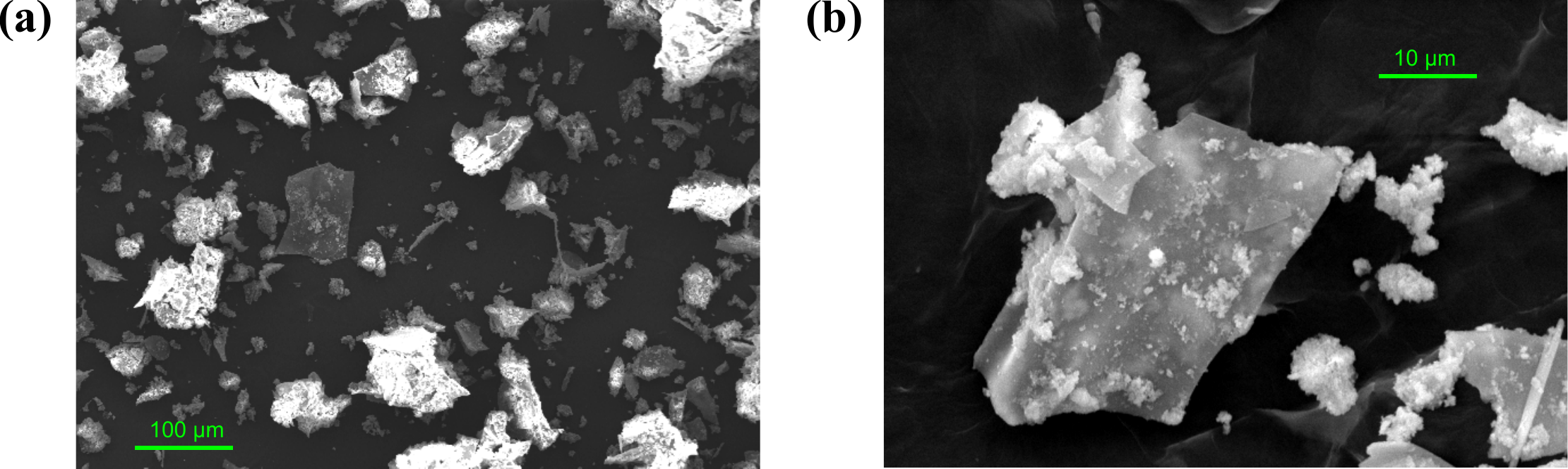}
\caption{Scanning Electron Microscopy images at (a)~$\times 500$ and (b)~$\times 5000$ magnification of the sludge after rinsing and drying. In (b), the large gray particle is constituted of perlite, while the smaller, whiter aggregates are barium sulfate.}\label{fig1}
\end{figure}

Table~\ref{tab:prop} gathers a few macroscopic physico-chemical properties of our surrogate sludge. The full list of constituents can be found in Appendix~\ref{secA1}. In brief, the sludge composition can be separated between (\textit{i})~a {\it soluble} ionic fraction composed mainly of Na$^{+}$, NO$_{3}^{-}$ and SO$_{4}^{2-}$ in water, which results in an ionic strength of 0.5 mol.L$^{-1}$, and (\textit{ii})~an {\it insoluble} fraction containing both colloidal and non-colloidal particles. The colloidal particles consist mainly of ferrous and cuprous hydroxides, and of the ferrocyanide complex Fe(CN)$_{6}$K$_{2}$Ni. These particles undergo attractive interactions due to the screening of electrostatic repulsion that results from the high ionic strength of the suspending aqueous phase. The non-colloidal particles are mainly barium sulfate BaSO$_{4}$ particles with a mean size of about 3~$\mu$m, and perlite particles from the filtration medium, with a mean size of about 30~$\mu$m. These large particles are widely polydisperse and non-spherical, with sizes ranging up to about 100~$\mu$m. Figure~\ref{fig1} shows electron microscopy images of the sludge after rinsing and drying.

Overall, the present sludge can be considered as a dispersion of non-Brownian particles into an attractive colloidal dispersion. The total solid volume fraction is 9.6~\%. It is important to note that the colloidal particles form a gel-like, percolated matrix, whose yield stress is sufficient for the non-Brownian particles to remain suspended without any significant sedimentation over several days. Prior to any measurement, the bottle containing the sludge is vigorously shaken by hand in order to fully fluidize and resuspend the whole sample.


\subsection{Rheology coupled to ultrasound imaging}\label{subsec22}

\subsubsection{Rheological measurements}\label{subsubsec221}

Rheological measurements are carried out using a stress-imposed rheometer (TA Instruments ARG2) equipped with a custom-made concentric-cylinder geometry made out of poly(methyl methacrylate) (PMMA). The radius of the rotating bob is $R_{1} = 23$~mm  and the radius of the cup is $R_{2} = 25$~mm, leaving a gap of $e = R_{2}-R_{1} = 2$~mm between the two cylinders. The height of the bob is $H = 60 \, \text{mm}$ and its bottom surface is flat. The distance between the bob and the bottom of the cup was set to 1~mm. The surfaces of both the bob and the cup are sandblasted, yielding a roughness of a few microns. As we shall see below, upon shearing, the sludge under study shows large slip velocities at both walls, in spite of significant surface roughness. This requires that rheometry be complemented with local velocity measurements in order to quantify wall slip and correct for its effect.

In this paper, we shall either impose the shear stress $\sigma$ or the shear rate $\dot\gamma$ by using the feedback loop of the rheometer. Note that the bob rotation speed $\Omega$ is linked to $\dot\gamma$ through:
\begin{equation}
\dot\gamma =\frac{R_1^2+R_2^2}{R_2^2-R_1^2}\,\Omega\,.
\label{eq:gpt}
\end{equation}
This results from the definition of the shear rate $\dot\gamma$ indicated by the rheometer, also referred to as the ``engineering'' shear rate, as the average shear rate in the case of a Newtonian fluid. In the small-gap approximation $e\ll R_1$, this equation leads to $\Omega\simeq \dot\gamma e/R_1$, so that the velocity at the bob periphery reads $v_0=R_1\Omega\simeq \dot\gamma e$, as expected for simple plane shear. More generally, with Eq.~\eqref{eq:gpt}, one has $v_0=\alpha\dot\gamma e$, where the dimensionless coefficient $\alpha$ is given by:
\begin{equation}
\alpha =\frac{R_2^2-R_1^2}{R_1^2+R_2^2}\,\frac{R_1}{e}\,.
\label{eq:alpha}
\end{equation}
The present concentric-cylinder geometry corresponds to $\alpha\simeq 0.957$.

Finally, the ``engineering'' shear stress $\sigma$ indicated by the rheometer is derived from the torque $\Gamma$ applied to the bob through:
\begin{equation}
\sigma =\frac{R_1^2+R_2^2}{4\pi H R_1^2R_2^2}\,\Gamma\,,
\label{eq:sigma}
\end{equation}
which again corresponds to the shear stress averaged across the gap in the case of a Newtonian fluid. Due to the curvature of the concentric-cylindrical geometry, the local shear stress $\sigma(r)$ actually decreases with the radial distance $r$ to the bob as $\sigma(r)=\Gamma/[2\pi H(R_1+r)^2]$. In the present geometry, this leads to a decrease of the shear stress from $\sigma(r=0)=\Gamma/(2\pi H R_1^2)$ at the bob to $\sigma(r=e)=\Gamma/(2\pi H R_2^2)$ at the cup, corresponding to a relative variation $\delta\sigma/\sigma=[\sigma(0)-\sigma(e)]/\sigma\simeq 2e/R_1\simeq 17\%$.

\subsubsection{Set-up for ultrasound imaging under shear}\label{subsubsec222}

We investigate the flow properties of our surrogate sludge thanks to ultrasound imaging coupled to standard rheological measurements. This combined technique provides access to local strain and velocity fields, simultaneously with the global rheological data, within optically opaque materials sheared in a concentric-cylinder geometry \cite{Manneville:2004a,Gallot:2013}. More precisely, our ultrasound technique is based on the analysis of the scattering of plane ultrasound pulses propagating through the material under study, with a small but non-zero angle relative to the radial direction of the concentric-cylinder geometry. In the present sludge, ultrasound scattering results from the presence of the non-Brownian suspended particles. Ultrasound images of the medium are reconstructed using a standard parallel beamforming algorithm \cite{Gallot:2013}. By repeating the acquisition of such images over time and by cross-correlating successive images, the projection $v_y$ of the velocity vector along the ultrasound propagation axis $y$ is measured. Provided the flow remains purely azimuthal, an assumption that is taken to be valid for the present yield stress material, the projection $v_y$ yields the time-resolved tangential component $v_\theta(r,z,t)$ of the velocity field in cylindrical coordinates $(r,z)$, where $r$ denotes the distance from the rotating bob, $z$ the vertical position measured from the bottom of the ultrasonic probe, and $t$ the time. The reader is referred to \cite{Gallot:2013} for full details on the tracking algorithm and on the calibration procedure used to recover $v_\theta(r,z,t)$ from $v_y$.

\begin{figure}[t]%
\centering
\includegraphics[width=0.8\textwidth]{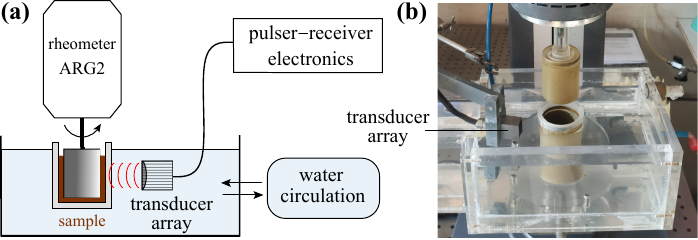}
\caption{(a) Sketch of the experimental set-up for ultrasound imaging coupled to rheometry. (b)~Picture of the experimental set-up after shearing the sludge (brown sample) and lifting up the bob.}\label{fig2}
\end{figure}

Figure~\ref{fig2} shows a sketch and a picture of the experimental set-up. The shear cell is immersed into a large water bath, whose temperature is kept constant at $25\pm0.1~^\circ$C and which allows for the transmission of ultrasound from/to the ultrasonic probe. The ultrasonic probe consists of an array of 128 independent channels stacked along the vertical direction and spaced by 50~$\mu$m. Each channel works both in emission and transmission, and is made of a piezoelectric transducer of width 200~$\mu$m. The transducer array is connected to a custom-made ultrasound scanner (Rack OPEN 128, Lecoeur Electronique), which emits pulses with a central frequency of $f = 15$~MHz and records the signal that is backscattered along the pulse propagation by the non-Brownian particles within the sheared sample. The region of interest thus corresponds to a slice of the sample of height 32~mm in the vertical direction $z$ and depth $e=2$~mm in the radial direction $r$. In the following, the origin $z=0$, corresponding to the bottom of the transducer array, is located about 15~mm above the bottom of the bob measuring position, such that the region of interest is centered around the middle height of the bob. 

\subsubsection{Analysis of velocity data}\label{subsubsec223}

At every point in time ${t}_{0}$ and for each height $z_0$, we estimate the fluid velocity $v_1$ at the bob surface by extrapolating the velocity profile $v_\theta(r,z_0,t_0)$ to $r=0$ thanks to a linear fit as a function of $r$. The same method is used to extract the fluid velocity $v_2$ at the cup surface, i.e., at $r=e=2$~mm. Linear fits are performed over 0.6~mm close to the walls. The point closest to the bob is left out of the analysis, since velocities  very close to a moving wall may be underestimated due to the combination of slight imperfections in the bob cylindrical shape and the procedure used for echo cancellation \cite{Gallot:2013}. Based on $v_1$ and $v_2$, we deduce the relative slip velocities at both walls, respectively $V_\text{bob}=(v_0-v_1)/v_0$ at the bob and $V_\text{cup}=v_2/v_0$ at the cup, where $v_0$ is the velocity of the bob surface as already defined above in Sect.~\ref{subsubsec221}. 

Based on the fluid velocities $v_1$ and $v_2$ in the vicinity of the shearing surfaces, we further define the {\it effective} shear rate ${\dot\gamma}_\mathrm{eff}$ by:
\begin{equation}
\dot\gamma_\mathrm{eff} = \frac{1}{\alpha}\,\frac{{v}_1-{v}_2}{e}\label{eq1}\,,
\end{equation}
where the dimensionless factor $\alpha$ allows for a direct comparison with the engineering shear rate $\dot\gamma$. In particular, in the presence of wall slip but when the bulk flow remains homogeneously sheared, ${\dot\gamma}_\mathrm{eff}$ is the actual shear rate experienced by the bulk material. Therefore, plotting the shear stress as a function of ${\dot\gamma}_\mathrm{eff}$ instead of ${\dot\gamma}$ allows us to correct for the presence of wall slip in the flow curve.

Finally, since we shall deal with time-dependent heterogeneous flows, it is important to note that $v_1$, $v_2$, $V_\text{bob}$, $V_\text{cup}$, and $\dot\gamma_\mathrm{eff}$ may actually depend both on the time $t_0$ at which the velocity map is acquired and on the vertical position $z_0$ along the transducer array. In the following, we shall first focus on these observables averaged on both $t_0$ and $z_0$, while keeping the same notations for simplicity. As for the velocity field $v_\theta(r,z,t)$, we shall simply note $v(r)=\langle v_\theta(r,z,t)\rangle_{z,t}$ its time- and $z$-average, $v(r,t)=\langle v_\theta(r,z,t)\rangle_{z}$ its time-resolved $z$-average, and $v(z,t)=\langle v_\theta(r,z,t)\rangle_{r}$ its time-resolved $r$-average. 

\subsection{Rheological protocols}\label{subsec3}

\subsubsection{Shear rate imposed tests}\label{subsubsec2}

A freshly fluidized sludge sample is loaded into the shear cell and subjected to an initial preshear at $\dot\gamma = 200 \, \text{s}^{-1}$ for 1~min. The shear rate is then logarithmically swept down from $200 \, \text{s}^{-1}$ to $0.2 \, \text{s}^{-1}$ within 150~s, and then increased back up to $200 \, \text{s}^{-1}$ with the same sweep rate. This flow curve is used as a reference to make sure that the initial state is reproducible from one loading to the other. The sample then undergoes a series of steps during which the shear rate is kept constant for 200~s with the following decreasing values: $\dot\gamma=200$, 100, 75, 40, 30, 20, 15, 10, 8, 5, 3, 2, 1, 0.5 and $0.2 \, \text{s}^{-1}$. For each step, ultrasound flow imaging is turned on after 60~s of shearing, and velocity maps $v_\theta(r,z,t)$ are acquired over a duration of 100~s. The number of velocity maps for each step depends on the shear rate: it is fixed to 100 maps acquired every second for $\dot\gamma \ge 2 \, \text{s}^{-1}$, while it decreases with the applied shear rate for $\dot\gamma \le 1 \, \text{s}^{-1}$, down to 15 maps recorded every 7~s at $\dot\gamma=0.2 \, \text{s}^{-1}$.

\subsubsection{Creep tests}
\label{subsubsec3}

Similar to the previous protocol, a preshear step at $\dot\gamma = 20 \, \text{s}^{-1}$ is applied to a fresh sample during 3 min, after which a downward sweep in shear rate is applied between $\dot\gamma = 20 \, \text{s}^{-1}$ and $0.2 \, \text{s}^{-1}$ within 1 min, followed by the reverse upward sweep. The sample is then left to rest at $\sigma = 0 \, \text{Pa}$ for 5 min. During this rest period, the sludge rebuilds a solid-like behaviour. A constant stress $\sigma$ is then applied to the sample and the resulting shear rate $\dot\gamma(t)$ is monitored for at least 300~s. Ultrasound imaging is started about 10~s before the start of the creep test and is performed over 200~s with one velocity map per second.

\section{Results}\label{sec3}

\subsection{Shear rate imposed tests}\label{subsec1}

\begin{figure}[hbt]%
\centering
\includegraphics[width=0.8\textwidth]{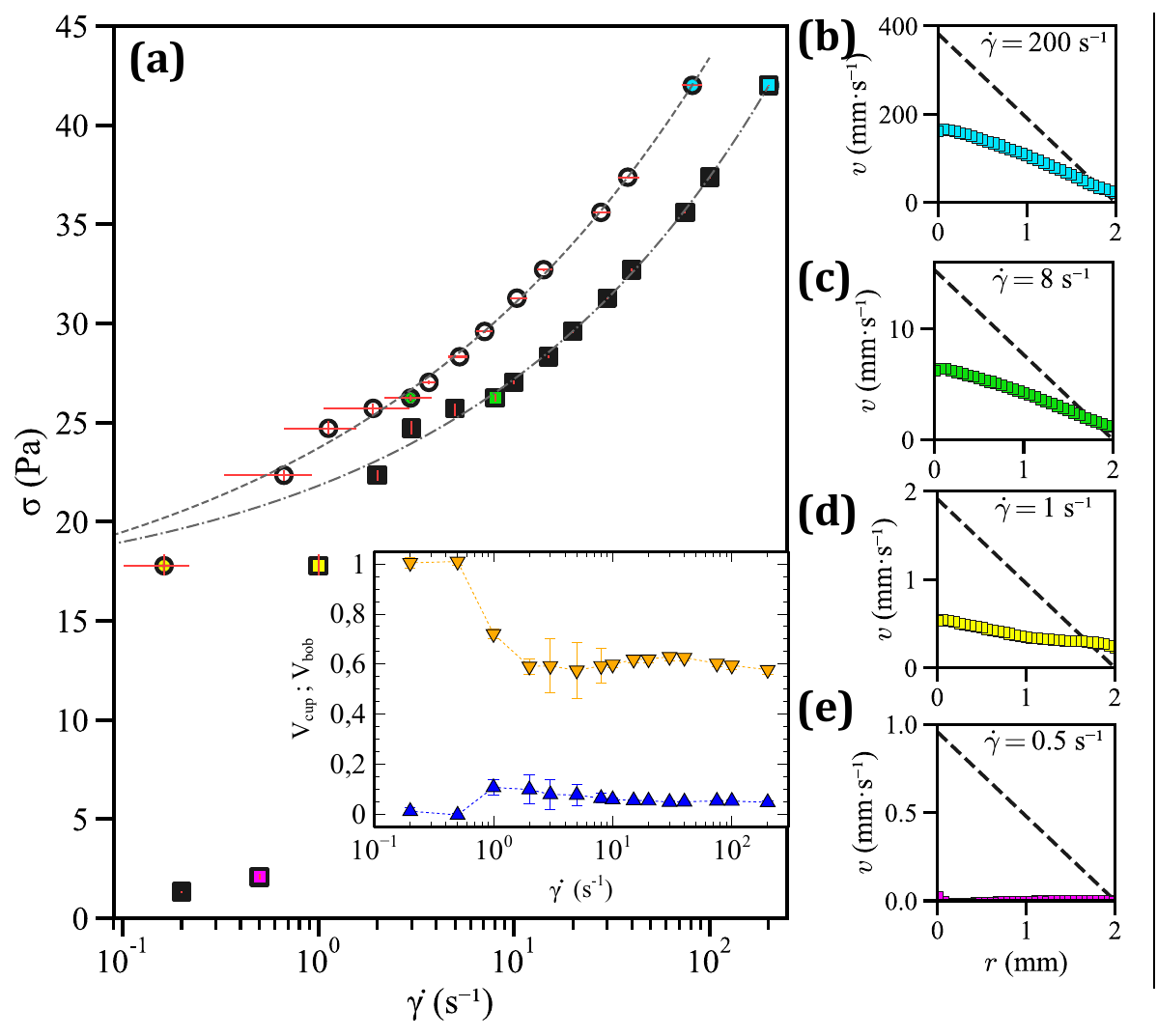}
\caption{(a)~Flow curve obtained through a series of steps where a constant shear rate is imposed for 200~s with values decreasing from $200 \, \text{s}^{-1}$ to $0.2 \, \text{s}^{-1}$. Rheological data are averaged over the last 50~s of each interval. Filled squares correspond to the ``engineering'' data $\sigma$ vs $\dot\gamma$ as measured by the rheometer, while empty symbols show $\sigma$ plotted as a function of the effective shear rate ${\dot\gamma}_\mathrm{eff}$ derived from the local velocity maps [see Eq.~\eqref{eq1} in Sect.~\ref{subsubsec223}]. Red lines indicate the range of variation of the various observables over the last 50~s of each interval. The gray dashed and dash-dotted lines show the best Herschel-Bulkley fits of the flow curves [see Eq.~\eqref{eq:HB} and Table~\ref{tab1}]. Inset: relative slip velocities at the bob $V_\text{bob}$($\triangledown$) and at the cup $V_\text{cup}$ ($\triangle$) as a function of $\dot\gamma$. The data are computed from the local velocity maps as described in Sect.~\ref{subsubsec223} and averaged over the last 50~s of each step. Error bars correspond to the standard deviation over the last 50~s of each step. (b)--(e) Velocity profiles $v(r)$ obtained from velocity maps $v_\theta(r,z,t)$ averaged over the vertical direction $z$ and over the last 50~s of, respectively, the step at (b)~${\dot\gamma} = 200 \, \text{s}^{-1}$, (c)~${\dot\gamma} =8 \, \text{s}^{-1}$, (d)~${\dot\gamma} =1 \, \text{s}^{-1}$, and (e)~${\dot\gamma} =0.5 \, \text{s}^{-1}$. The bob surface is located at $r=0 \, \text{mm}$, while the cup wall is located at $r=2 \, \text{mm}$. Dashed lines show the linear velocity profile expected for a Newtonian fluid in the absence of wall slip. The points corresponding to (b), (c), (d), and (e) in the flow curves are highlighted, respectively, by blue, green, yellow and magenta symbols in (a).}\label{fig3}
\end{figure}

Figure~\ref{fig3} summarizes the results obtained through series of steps under imposed shear rates as described above in Sect.~\ref{subsubsec2}. In order to focus on the steady-state reached at the end of each step, both rheological and velocimetry data are averaged over the last 50~s of each interval. Filled symbols in Fig.~\ref{fig3}(a) correspond to the engineering flow curve, $\sigma$ vs $\dot\gamma$. The time- and $z$-averaged velocity profiles $v(r)$ recorded for $\dot\gamma=200 \, \text{s}^{-1}$, $8 \, \text{s}^{-1}$, $1 \, \text{s}^{-1}$, and $0.5 \, \text{s}^{-1}$ are displayed in Fig.~\ref{fig3}(b), (c), (d), and (e) respectively. All
velocity profiles show strong wall slip at the bob surface $r=0$, where the fluid velocity is always at least twice as low as the bob velocity $v_0$. Wall slip is also noticeable at the cup surface $r=e$. Similar observations were made for all shear rates in the range under study, with a significant increase of relative slip velocities for the lower shear rates [see inset in Fig.~\ref{fig3}(a)], and in spite of the rather large surface roughness obtained through sandblasting at both walls.

For $\dot\gamma\gtrsim 1 \, \text{s}^{-1}$, the bulk of the sludge undergoes homogeneous shear, i.e., the velocity profile is linear with a non-zero effective shear rate $\dot\gamma_\mathrm{eff}$ [see Fig.~\ref{fig3}(b,c)]. However, below a certain critical shear rate, here around $1 \,\text{s}^{-1}$, the flow completely stops and reaches a regime of total wall slip with effective shear rate ${\dot\gamma}_\mathrm{eff}=0$ [see Fig.~\ref{fig3}(d,e)]. This transition corresponds to a sharp stress drop in the flow curve [see yellow and magenta points in Fig.~\ref{fig3}(a)]. Such a transition to a plug-like flow at low shear rates has been reported many times for yield stress fluids in the literature \cite{Barnes:1995, Bertola:2003, Meeker:2004, Seth:2012, Zhang:2017, Cloitre:2017, Liberto:2020}. However, in the present case, strong wall slip is reported all along the flow curve. This prompts us to plot the ``true'' flow curve $\sigma$ vs $\dot\gamma_\mathrm{eff}$ [see empty symbols in Fig.~\ref{fig3}(a)]. Correcting the global rheological data for wall slip thus leads to a significant shift of the flow curve towards lower shear rates. As a consequence, fitting both flow curves by the empirical Herschel-Bulkley model,
\begin{equation}
\sigma = \sigma_0 + K \dot\gamma^n\,,
\label{eq:HB}
\end{equation}
leads to significantly different estimates of the yield stress $\sigma_0$, the consistency index $K$, and the power-law exponent $n$ (see Table~\ref{tab1}).

\begin{table}[htb]
\caption{Yield stress $\sigma_0$, consistency index $K$, and power-law exponent $n$ inferred from fits of the flow curves in Fig.~\ref{fig3} with the Herschel-Bulkley model [see Eq.~\eqref{eq:HB}]. The last column indicates the range of shear rates used in the fits.}\label{tab1}%
\begin{tabular}{@{}ccccc@{}}
\toprule
Data set & $\sigma_{0} \, \text{(Pa)}$ & $K \, \text{(Pa.s}^n\text{)}$ & $n$ & Fitting range\\
\midrule
$\sigma$ vs ${\dot\gamma}$    & 15.8 & 6.0 & 0.28 & 2--200~s$^{-1}$\\
$\sigma$ vs ${\dot\gamma}_\mathrm{eff}$    & 13.6 & 10.1 & 0.23 & 0.5--200~s$^{-1}$\\
\botrule
\end{tabular}
\end{table}

\begin{figure}[h]%
\centering
\includegraphics[width=1\textwidth]{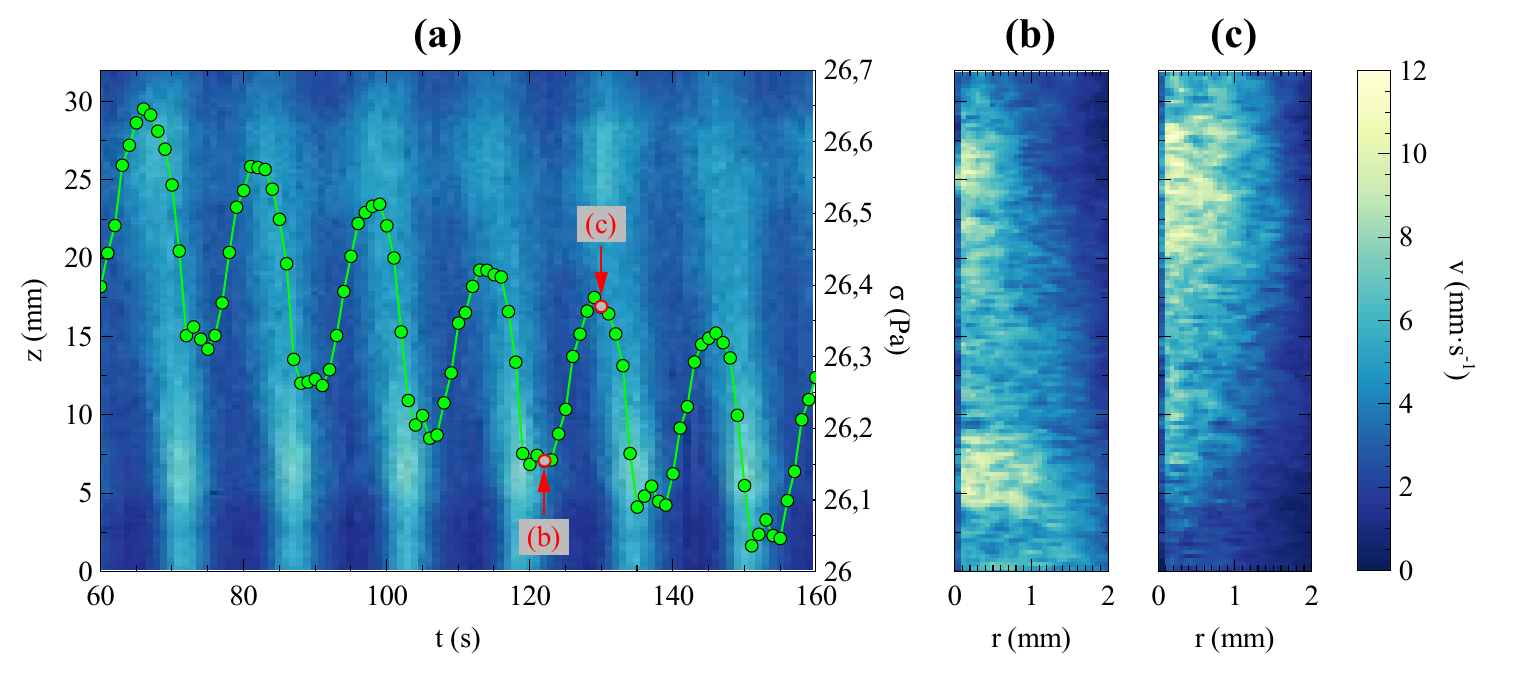}
\caption{(a)~Spatiotemporal diagram of the velocity field ${v}(z,t)$ averaged over the radial direction $r$ and shown as a function of time $t$ and vertical position $z$. The imposed shear rate is $\dot\gamma = 8 \, \text{s}^{-1}$, and ultrasound imaging is started 60~s after the beginning of the shear rate step. The velocity is coded using a linear scale as shown in the color bar on the right. Green dots show the shear stress response $\sigma(t)$ (left axis) recorded by the rheometer simultaneously to ultrasound imaging. (b,c)~Velocity map $v_\theta(r,z,t)$ recorded respectively at $t=122 \, \text{s}$ and $t=130 \, \text{s}$. See also Supplemental Movie~1.}\label{fig4}
\end{figure}

\begin{figure}[t]%
\centering
\includegraphics[width=0.45\textwidth]{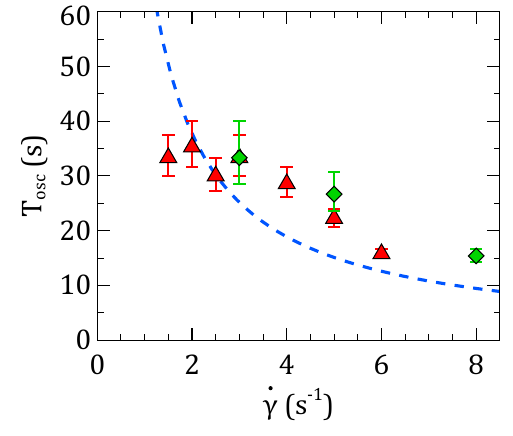}
\caption{Characteristic period $T_\text{osc}$ of the stress oscillations (green diamonds) recorded for 1~s$^{-1}\lesssim\dot\gamma\lesssim 8$~s$^{-1}$. The red triangles correspond to a rheological protocol that slightly differs from that described in Sect.~\ref{subsubsec2} and where a preshear is applied at 20~s$^{-1}$ for 600~s followed by steps at $\dot\gamma=10$, 9, 8, 7, 6, 5, 4, 3, 2.5, 2, 1.5, 1, and 0.5~$\, \text{s}^{-1}$ applied for 300~s each. The blue dashed line shows the period of rotation of the bob $T_\text{bob}$ as a function of $\dot\gamma$.}\label{fig5}
\end{figure}

Interestingly, we point out that, for shear rates ranging between about $1  \, \text{s}^{-1}$ and $8 \, \text{s}^{-1}$, the flow is noticeably heterogeneous both in time and space. Such heterogeneity is suggested by the large temporal fluctuations of ${\dot\gamma}_\mathrm{eff}$, as indicated by the red horizontal lines in Fig.~\ref{fig3}(a) and by the large standard deviation of the relative slip velocities in the inset of Fig.~\ref{fig3}(a). Figure~\ref{fig4} further focuses on the local velocity field recorded over time at $\dot\gamma = 8 \, \text{s}^{-1}$. It is clear from Fig.~\ref{fig4}(a) that the flow undergoes spontaneous oscillations with a period of $T_\text{osc}=16.7\pm 1$~s, which differs significantly from the rotation period of the bob $T_\text{bob}=2\pi/\Omega\simeq 9.5\, \text{s}$ for $\dot\gamma = 8 \, \text{s}^{-1}$. These fluctuations are also reflected in the shear stress $\sigma(t)$ recorded by the rheometer, which shows small-amplitude pseudo-periodic oscillations that are synchronized with those of the velocity field [see green dots in Fig.~\ref{fig4}(a)]. Looking more closely at the velocity field, it appears that, while the sludge is homogeneously sheared in the radial direction, the flow is spatially heterogeneous along the vertical direction $z$.  
Indeed, velocity fluctuations propagate from the top to the bottom of the shear cell. Two typical snapshots of $v_\theta(r,z,t)$ are shown in Fig.~\ref{fig4}(b,c) for the two times indicated by red arrows in Fig.~\ref{fig4}(a). As also shown in Supplemental Movie~1, the velocity fluctuations consist in strong variations of the slip velocity at the bob surface, which are initiated periodically at the bottom of the cell and propagate upwards. Finally, Fig.~\ref{fig5} shows that, for the range of shear rates where fluctuations are reported, the oscillation period $T_\text{osc}$ of the stress signal is of the order of, but not strictly equal to, the bob rotation period $T_\text{bob}$.

\begin{figure}[t]%
\centering
\includegraphics[width=0.9\textwidth]{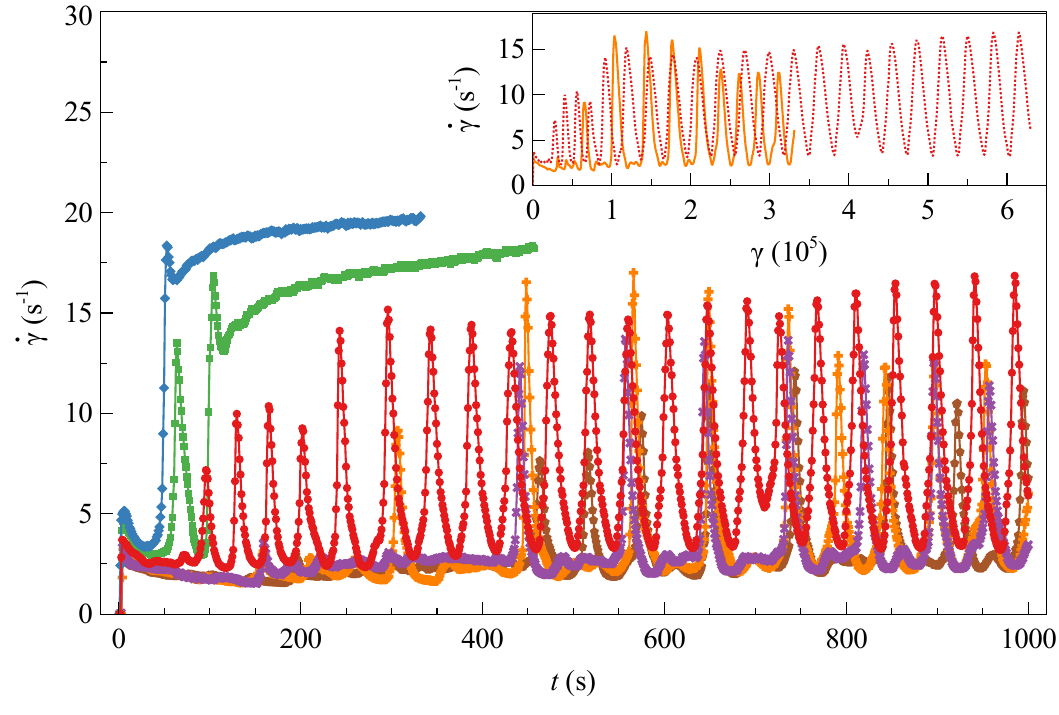}
\caption{Shear rate $\dot\gamma$ recorded as a function of  time $t$ under a constant shear stress $\sigma$ applied from $t=0$, with $\sigma=29.0$~Pa (brown), 30.0~Pa (purple), 30.5~Pa (orange), 30.8~Pa (red), 30.9~Pa (green) and 31.0~Pa (blue). Data are acquired with one point per second, and lines between markers are drawn for visual aid. Inset: shear rate $\dot\gamma$ recorded for $\sigma=30.5$~Pa (orange) and 30.8~Pa (red), and plotted as a function of the accumulated strain $\gamma(t)=\int_0^t \dot\gamma(t')\,\mathrm{d}t'$. See also Supplemental Movies~2, 3, and 4 for velocity maps $v_\theta(r,z,t)$ recorded during the first 200~s of the creep tests at $\sigma=30.5$~Pa, 30.8~Pa, and 30.9~Pa respectively.}\label{fig6}
\end{figure}

\subsection{Stress imposed tests}\label{subsec2}

To get more insight into the unusual flow behavior of the sludge under study, we turn to creep tests, where a constant stress $\sigma$ is applied following the protocol described in Sect.~\ref{subsubsec3}, and ultrasound imaging is performed simultaneously. Figure~\ref{fig6} gathers the shear rate responses of solid-like sludge samples subject to shear stresses close to the yield stress, ranging from 29~Pa to 31~Pa. Note that this range of stresses lies significantly above the yield stress values reported in Table~\ref{tab1}. This is because measurements from a fluidized state, as performed above in Sect.~\ref{subsec1}, probe the {\it dynamic} stress, which is smaller than the {\it static} yield stress probed here from a solid-like state. 

For the two lower stresses, $\sigma=29$ and 30~Pa, the shear rate recorded by the rheometer remains initially small, at about 2--$3 \, \mathrm{s}^{-1}$, for about 400~s. After this induction period, the shear rate occasionally shows sharp increases up to 10--$15 \, \mathrm{s}^{-1}$, before returning to the initial ``low-shear'' state. Such peaks last about $30$~s each and are observed until the end of the test. As the applied stress is increased to $\sigma=30.5$~Pa and 30.8~Pa, this striking oscillatory behavior starts at earlier times and occurs more frequently and in a more regular manner. When $\dot\gamma(t)$ is plotted over time, the oscillations are far from sinusoidal, which suggests strong nonlinearity. When $\dot\gamma(t)$ is plotted as a function of the accumulated strain $\gamma=\int_0^t \dot\gamma(t')\,\mathrm{d}t'$, the oscillations appear much more sinusoidal, but their period in strain units depends on the applied stress (see inset in Fig.~\ref{fig6}). Thus, the phenomenon is not simply driven by the bob position and rotation period. Finally, for $\sigma=30.9$~Pa and 31~Pa, the shear rate reaches a ``high-shear'' steady state within a few minutes, possibly after a couple of peaks. 

\begin{figure}[h]%
\centering
\includegraphics[width=1\textwidth]{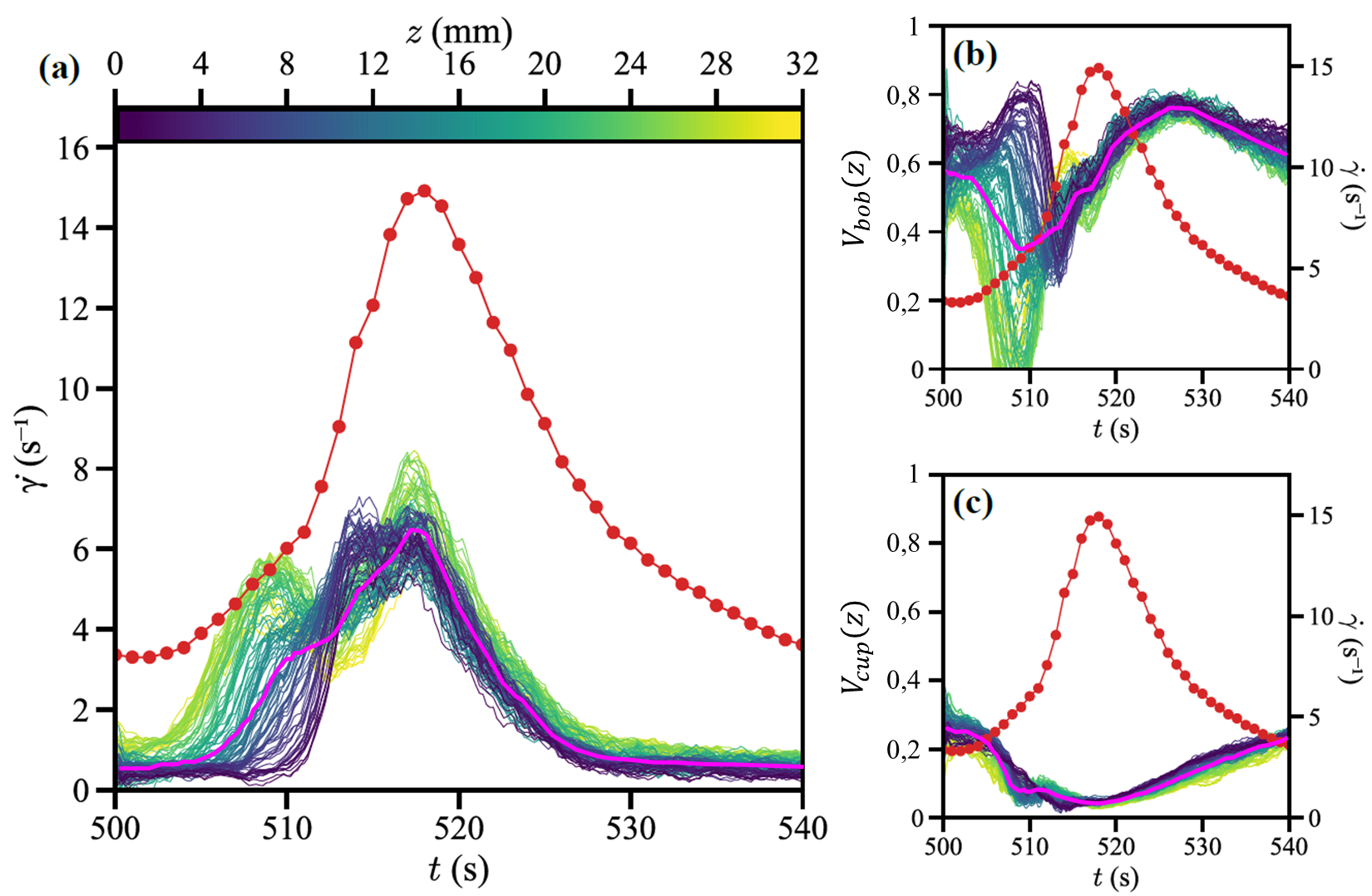}
\caption{(a)~Local effective shear rate $\dot\gamma_\mathrm{eff}(z,t)$ and (b,c)~relative slip velocities $V_\text{bob}(z,t)$ and $V_\text{cup}(z,t)$, respectively, at the bob and at the cup, recorded as a function of time $t$ under a constant applied shear stress $\sigma = 30.8$~Pa. Effective shear rates and relative slip velocities are computed as described in Sect.~\ref{subsubsec223} for each position $z$ along the ultrasonic probe. The purple line shows the average over $z$ of effective shear rates and relative slip velocities. The data were smoothed by using a moving average over 2~s. Colors code for the vertical position $z$ from dark blue at the bottom to yellow at the top of the region of interest. The data in red correspond to the engineering shear rate $\dot\gamma(t)$ measured by the rheometer [see right axes in (b) and (c)]. See also Supplemental Movie~5 for the velocity maps $v_\theta(r,z,t)$.}
\label{fig7}
\end{figure}

In order to draw a local picture of the remarkable dynamical behavior revealed in Fig.~\ref{fig6}, ultrasound imaging was performed as described in Sect.~\ref{subsubsec3}. Movies of the velocity maps $v_\theta(r,z,t)$ are provided in the Electronic Supplemental Information. Supplemental Movie~2 shows that, for $\sigma=30.5$~Pa, the sample undergoes a plug-like flow with total wall slip for the whole 200~s duration of the ultrasound acquisition, which corresponds roughly to the first three minutes of the creep test. Similar results are obtained for $\sigma=29$ and 30~Pa. We conclude that the ``low-shear'' state, where $\dot\gamma\simeq 2$--$3 \, \mathrm{s}^{-1}$, actually corresponds to a solid-like state where the sample remains unsheared, i.e., $\dot\gamma_\mathrm{eff}=0$, and all the shear deformation is localized in lubricating layers at the walls. For $\sigma=30.8$~Pa, however, Supplemental Movie~3 indicates that the successive peaks in $\dot\gamma(t)$ are associated with complex spatiotemporal flow dynamics where $v_\theta(r,z,t)$ abruptly increases and the bulk sample is sheared, yet with strong variations along the $z$ direction. Here, the frame rate of only 1~Hz is not sufficient to fully resolve the local flow dynamics, which are further explored below thanks to faster imaging. Finally, for $\sigma=30.9$~Pa, Supplemental Movie~4 shows that the asymptotic ``high-shear'' state, reached after a single peak event, corresponds to a liquid-like state where the sludge is homogeneously sheared throughout the whole region of interest, although a large amount of wall slip persists in the steady state. 

Supplemental Movie~5 provides deeper insight into the local scenario associated with the peaks in shear rate thanks to a high time resolution. Here, 200~ultrasound images were recorded over 40~s with a frame rate of 5~fps during the peak event that occurs from $t=500$~s to $t=540$~s at $\sigma=30.8$~Pa. The analysis of the velocity maps $v_\theta(r,z,t)$ is displayed in Fig.~\ref{fig7}. Figure~\ref{fig7}(a) compares the engineering shear rate $\dot\gamma(t)$ measured by the rheometer (in red) to the local effective shear rates $\dot\gamma_\mathrm{eff}(z,t)$ defined following the procedure described in Sect.~\ref{subsubsec223} at each vertical position $z$ along the ultrasonic probe and coded from dark blue at the bottom to yellow at the top of the region of interest. The corresponding relative slip velocities at the bob and at the cup, $V_\text{bob}(z,t)$ and $V_\text{cup}(z,t)$, are shown, respectively, in Figs.~\ref{fig7}(b) and (c) using the same color code. Before and after the shear rate peak, the local shear rate in the bulk vanishes, which corresponds to plug-like flow with relative slip velocities of about 70\% at the bob and 30\% at the cup. At the beginning of the recording, the local shear rate at the top of the region of interest starts to increase up to the point where it reaches the ``engineering'' shear rate $\dot\gamma$ for $t= 507$--510~s [see yellow to green curves in Fig.~\ref{fig7}(a)]. This indicates that, at this specific location along the vertical direction, the sample fully {\it sticks} at the walls, as confirmed by the vanishingly small slip velocities in Figs.~\ref{fig7}(b) and (c). Meanwhile, the temporal increase in $\dot\gamma_\mathrm{eff}(z,t)$, associated with a decrease in $V_\text{bob}(z,t)$ {\it propagates} from the top to the bottom of the cell at about 4~mm.s$^{-1}$ until a homogeneous shear flow is reached around $t\simeq 515$~s with $\dot\gamma_\mathrm{eff}\simeq 6$~s$^{-1}$ and about 50\% of wall slip, whatever the vertical position $z$. The flow then remains $z$-independent for the rest of the event. Still, $\dot\gamma_\mathrm{eff}$ strongly decreases and effectively vanishes at $t\simeq 530$~s, while the relative slip velocity increases from about 50\% to 80\% at the bob and from about 5\% to 20\% at the cup. Then, for $t\gtrsim 530$~s, the local effective shear rate remains negligible, and the relative slip velocities go back to their values of about 70\% at the bob and 30\% at the cup, corresponding to the fully arrested state reported prior to the peak event. Overall, we conclude that one cycle in the shear rate oscillations observed in Fig.~\ref{fig6} corresponds to complex local stick-slip dynamics that involve the propagation of a ``stick event'' followed by the emergence of a homogeneously sheared flow around the maximum in shear rate, which progressively gives way to a plug-like flow.

\section{Discussion and conclusion}\label{sec4}


We have shown that a complex suspension representative of some sludges produced by the nuclear industry exhibits oscillating dynamics in the vicinity of the yield stress, both under an imposed shear rate and under an imposed shear stress. This phenomenology, revealed by ultrasonic imaging coupled with rheometry, is reminiscent of stick-slip instabilities or, more generally, bistability in dynamical systems \cite{Salmon:2002}. Here, close to the shear-induced solid-to-liquid transition and under an imposed stress, the flow oscillates between an unsheared plug-like regime and a homogeneously sheared flow profile, i.e. between a solid-like state and a fluid-like state (see Figs.~\ref{fig6} and \ref{fig7}). Under an imposed shear rate (Fig.~\ref{fig4}), the phenomenology seems more complex, since the oscillating flow involves the propagation of less viscous regions (bearing a high local shear rate) within more viscous or even solid-like regions (experiencing a smaller shear rate or even no shear at all). Note that ultrasound imaging only provides access to a fixed two-dimensional slice across the Taylor-Couette cell, and that the heterogeneous flows described above may not have cylindrical symmetry. It is therefore conceivable that flow heterogeneities are localized along the azimuthal direction and/or propagate around the axis of rotation. However, the fact that oscillations are clearly detectable in the global rheological data suggests that the flow oscillates as a whole, and that the heterogeneities observed along the $z$-direction are present simultaneously over the entire circumference of the geometry. Still, it is clear that further analysis of slip dynamics, in particular by correlating fluctuations in the slip velocities and those in rheological quantities, is required to better understand their links with volume dynamics.

Dynamical phenomena similar to those described here have already been reported many times in the literature of complex fluids \cite{Wunenburger:2001,Manneville:2004b,Decruppe:2006,Ganapathy:2006,Manneville:2007,Gentile:2013,Skvortsov:2019}. In particular, since the pioneering results of \cite{Lootens:2003}, numerous studies have demonstrated the existence of high-amplitude, periodic or chaotic fluctuations in concentrated suspensions of both colloidal and granular particles in the vicinity of shear-thickening transitions or close to the jamming threshold \cite{Nagahiro:2013,Larsen:2014,Bossis:2017,Isa:2009,Malkin:2012,Hermes:2016,Rathee:2017,SaintMichel:2018,Ovarlez:2020,Maharjan:2021,Miller:2022}. ``Pulses'' associated with local variations in slip velocities and propagating along the direction of vorticity have thus been observed in dense cornstarch suspensions under an imposed stress \cite{SaintMichel:2018}. These dynamics, which are very similar to those reported here under an imposed shear rate (see Fig.~\ref{fig4}), have been interpreted as the signature of an instability associated with a non-monotonic flow curve \cite{Chacko:2018}. However, to the best of our knowledge, such time-dependent phenomena had not been observed so far in suspensions with a much more modest particle volume fraction (of the order of 12\% here vs. 40\% to over 50\% in dense suspensions).

In the surrogate sludge studied in this work, attractive interactions leading to a gel-like matrix are responsible for the yield stress and for the thixotropy of the sample. Although strong stress fluctuations have already been reported in colloidal gels in the vicinity of the yield stress or during the solid--liquid transition under an imposed shear rate \cite{Kurokawa:2015,Liberto:2020}, these were associated with shear-banding flows, where a significant fraction of the gap is sheared and coexists with a solid-like region. Here, no shear banding is reported and the oscillations shown in Fig.~\ref{fig6}, whose shape is reminiscent of the relaxation oscillations of non-linear oscillators as already reported in surfactant phases \cite{Salmon:2002}, are associated with a transition from a fully sheared flow to a plug-like flow. As already pointed out above, a detailed study of wall slip dynamics, particularly under an imposed stress, is needed to better understand the physics involved in these oscillations and to determine whether their origin lies in a stick-slip instability at the wall or in a volume instability. 
The possibility that gradients in the local volume fraction occur, or even density waves, as recently reported in dense shear-thickening suspensions \cite{Ovarlez:2020}, also calls for future microstructural investigations of the present sludge under shear.

\backmatter

\bmhead{Supplementary information}

Electronic Supplementary Information include five supplemental movies showing animations of the time-resolved velocity maps $v_\theta(r,z,t)$ inferred from ultrasound imaging.

\bmhead{Acknowledgments}

This work benefited from a Ph.D. grant funded by CEA under the collaboration agreement DES 6024/C42076.

\section*{Declarations}

Conflict of interest: the authors declare no conflict of interest.

\begin{appendices}

\section{Effluent treatment and ionic composition of the suspending liquid}\label{secA1}

The sludge was produced at a pilot scale by adding various reagents to a non-radioactive liquid effluent. This effluent contains different ions which concentrations reproduce the saline composition of typical radioactive effluents. The amount of the various reagents is based on the volume of the effluent to be treated as listed in Table~\ref{tabA1}).

\begin{table}[htbp]
\caption{Reagents and products of the liquid effluent treatment}\label{tabA1}%
\begin{tabular}{@{}ccc@{}}
\toprule
Reagent formula & Mass per liter of effluent (g) & Product type\\
\midrule
FeSO$_{4}\cdot 7$~H$_{2}$O  & 6.95 & Iron hydroxides \\
\midrule
CuSO$_{4} \cdot 5$~H$_{2}$O  & 1.39 & Cuprous hydroxides \\
\midrule
Na$_{2}$SO$_{4}$  & 5.56 & \multirow{2}*{BaSO$_{4}$}\\
Ba(NO$_{3}$)$_{2}$  & 1.67 & \\
\midrule
Fe(CN)$_{6}$K$_{4} \cdot 3$~H$_{2}$O  & 1.67 & \multirow{2}*{Fe(CN)$_{6}$NiK$_{2}$}\\
NiSO$_{4} \cdot 6$~H$_{2}$O  & 1.11 & \\
\botrule
\end{tabular}
\end{table}

In order to analyse the suspending liquid, the sludge was centrifuged at 4500 rpm for 15 minutes and the supernatant was retrieved and filtered using a PVDF syringe filter (0.2~$\mu$m). The supernatant composition was then determined by Inductively Coupled Plasma – Atomic Emission Spectroscopy (ICP-AES, Thermo Scientific iCAP 6000 Series) as listed in Table~\ref{tabA2}.

\begin{table}[htbp]
\caption{Composition of the suspending liquid as determined by ICP-AES.}\label{tabA2}%
\begin{tabular}{@{}cc@{}}
\toprule
Ion & Concentration ($mg\cdot L^{-1}$)\\
\midrule
$NO_{3}^{-}$  & 13 708 $\pm~1 371$\\
$SO_{4}^{2-}$  & 6 853 $\pm~685$\\
$Cl^{-}$  & 109 $\pm~11$\\
$Na^{+}$  & 8 535 $\pm~854$\\
$K^{+}$  & 381 $\pm~38$\\
$Ca^{2+}$  & 302 $\pm~30$\\
$Mg^{2+}$  & 174 $\pm~17$\\
\botrule
\end{tabular}
\end{table}

\end{appendices}



\pagebreak

\end{document}